\documentclass[twocolumn,epjc3]{svjour3}

\RequirePackage[T1]{fontenc}
\RequirePackage{mathptmx}     
\RequirePackage{latexsym}

\usepackage{amsmath}
\usepackage{amssymb}
\usepackage{float}

\usepackage{graphicx}
\usepackage{caption}
\usepackage{subcaption}
\usepackage{epstopdf}

\usepackage{flushend}
\usepackage{cuted}            
\usepackage{comment}
\usepackage[colorlinks=true,
            citecolor=blue,
            urlcolor=blue,
            linkcolor=blue]{hyperref}

\smartqed    

\journalname{Eur. Phys. J. C}

\begin{document}

\title{Exploration of Zero-Complexity Compact Stars in Higher Dimensions under the Finch-Skea Background}

\author{
Shyam Das\thanksref{e1,addr1}
\and
Megandhren Govender\thanksref{e2,addr2}
\and
Kevin Reddy\thanksref{e3,addr3}
\and
 Bikram Keshari Parida\thanksref{e4,addr4,addr5} 
}

\thankstext{e1}{e-mail: dasshyam321@gmail.com }
\thankstext{e2}{e-mail: megandhreng@dut.ac.za}
\thankstext{e3}{e-mail: kevinr@dut.ac.za}
\thankstext{e4}{e-mail: parida.bikram90.bkp@gmail.com}

\institute{
\label{addr1}
Department of Physics, Malda College, Malda, India
\and
\label{addr2}
Department of Mathematics, Faculty of Applied Sciences, Durban University of Technology, Durban 4000, South Africa 
\and
\label{addr3}
Department of Physics, Faculty of Applied Sciences, Durban University of Technology, Durban 4000, South Africa
\and
\label{addr4}
Department of Physics and Engineering Physics, Tulane University, New Orleans, Louisiana 70118, USA
\and\label{addr5}
Department of Data Science, Dongduk Women's University, Seoul, South Korea}

\maketitle   

\begin{abstract}
Motivated by the recent extension of Herrera's gravitational complexity to arbitrary higher-dimensional spacetimes, we investigate exact compact-star models that satisfy the vanishing-complexity condition within the Finch--Skea geometry in $(n+2)$-dimensional Einstein gravity. By combining the higher-dimensional Einstein field equations with the generalized complexity formalism, we obtain a new class of exact interior solutions describing anisotropic fluid spheres with zero complexity. The physical properties of these models are evaluated by analyzing the behavior of the matter variables, pressure anisotropy, and equilibrium conditions, alongside the standard requirements of regularity, energy conditions, causality, and stability. We further explore the influence of spacetime dimensionality on the structural characteristics of the stellar configurations, demonstrating that the presence of extra dimensions significantly modifies the internal matter distribution while preserving physical viability. The solutions presented here represent the first exact higher-dimensional Finch--Skea compact-star models constructed within this recently developed generalized complexity framework \cite{choudas}. These results provide a natural extension of zero-complexity stellar configurations beyond four-dimensional General Relativity and offer a robust framework for investigating self-gravitating systems in higher-
dimensional gravity.

\end{abstract}

\section{Introduction}
Compact stars are the dense remnants of massive stars, predicted by General Relativity. The intense gravitational fields and ultrahigh densities characteristic of these objects make them particularly useful for probing the behaviour of matter under extreme conditions. Remarkable progress in observational astronomy, including the discovery of massive neutron stars and the detection of gravitational waves from compact binary mergers, has significantly enhanced our understanding of dense matter and renewed interest in constructing realistic theoretical models of relativistic stellar objects \cite{tolman1939,oppenheimer1939,buchdahl1959,demorest2010,antoniadis2013}. Consequently, compact stars continue to provide an important bridge between gravitation, nuclear physics, particle physics, and astrophysics. Exact interior solutions to Einstein's field equations are fundamental to the study of relativistic compact stars. Since the foundational works of Schwarzschild \cite{schwarzschild1916}, Tolman \cite{tolman1939}, and Oppenheimer and Volkoff \cite{oppenheimer1939}, numerous analytical models have been proposed to investigate the structure and physical properties of compact objects. A physically acceptable compact-star model must possess regular metric functions, positive and well-behaved matter variables, satisfy the energy and causality conditions, remain stable, and match smoothly to the exterior spacetime at the stellar boundary \cite{delgaty1998,ivanov2002,mak2003,haensel2007}. Despite the mathematical complexity of the field equations, exact solutions remain valuable for understanding stellar equilibrium, stability, and relativistic effects.

Another important aspect of realistic compact-star modelling is the possible existence of local pressure anisotropy. Although the simplest stellar models assume an isotropic perfect fluid, several physical mechanisms--including phase transitions \cite{Sokolov1980}, superfluidity \cite{Sokolov1980,Canuto1974}, strong magnetic fields \cite{Weber1999}, viscosity \cite{Herrera2004}, pion condensation \cite{SawyerScalapino1973}, and solid cores \cite{CanutoChitre1973}--may generate unequal radial and tangential pressures inside compact objects. Ruderman \cite{ruderman1972} first pointed out that matter at supranuclear density may naturally exhibit anisotro-pic stresses, while Bowers and Laing \cite{bowers1974} demonstrated that pressure anisotropy can significantly influence the maximum mass, radius, redshift, and stability of relativistic stars. Herrera showed that dissipative fluxes, energy-density inhomogeneities, and  the presence of shear in a fluid flow inescapably drive an initially isotropic self-gravitating fluid distribution
to deviate from the isotropic-pressure condition, thus resulting in local pressure anisotropy \cite{Herrera2020}, an effect
expected to be crucially important during the late stages of stellar evolution \cite{HerreraSantos1997}. Thus, anisotropic fluid distributions have become an integral component of modern compact-star modeling and have led to the development of numerous physically acceptable stellar solutions having appreciable agreement with astrophysical observations.

Among the many exact interior solutions proposed in the literature, the Finch--Skea model \cite{finch1989} occupies a prominent position owing to its mathematical simplicity and remarkable physical behaviour. The regular behaviour of the metric functions, together with well-behaved thermodynamic variables, has made this geometry one of the most extensively investigated exact solutions in relativistic astrophysics. Over the years, the Finch--Skea ansatz has been successfully generalized to include anisotropic matter distributions, electrically charged compact stars, strange stars, embedding class-one spacetimes, and several modified theories of gravity, consistently yielding physically acceptable stellar configurations \cite{sharma2001,mak2004,murad2015,maurya2017,pant2019,maurya2019}. Owing to its geometric simplicity and physical viability, the Finch--Skea spacetime provides an effective framework for modelling compact stars. Accordingly, we adopt the Finch--Skea metric as one of the gravitational potentials in the present work.

Despite the considerable success achieved by exact stellar models, an important question remains regarding the internal structural organization of self-gravitating matter. Two compact stars may satisfy all conventional physical acceptability conditions \cite{delgaty1998} while exhibiting substantially different internal distributions of density and pressure \cite{herrera2018new}. Consequently, traditional analyses based solely on metric functions and matter variables may not be sufficient to characterize the intrinsic structural properties of relativistic fluid distributions. This observation naturally motivates the introduction of an additional physical quantity capable of measuring the internal organization of self-gravitating systems. In recent years, this idea has led to the development of the concept of \emph{gravitational complexity}, which has emerged as an important and rapidly growing area of research in relativistic astrophysics.

A pioneering contribution towards quantifying this ``complexity" was established by Herrera \cite{herrera2018new}, who proposed a scalar quantity---referred to as the complexity factor---to characterize the internal structure of static, spherically symmetric systems. This quantity effectively combines two key physical features of realistic matter distributions: the non-uniformity of the energy density and the anisotropy of the pressures. By doing so, it offers a coherent and quantitative way to assess how intricate a compact object is, with direct relevance to models of neutron stars and anisotropic stellar interiors, as well as to investigations of their equilibrium and stability.
However, Herrera's construction \cite{herrera2018new} was developed within the standard four-dimensional framework of General Relativity. Contemporary developments in gravitational theory, such as higher-dimensional extensions of Einstein gravity \cite{Kaluza1921,Klein1926,OverduinWesson1997}, braneworld scenarios \cite{RandallSundrum1999a,RandallSundrum1999b,MaartensKoyama2010}, and string-inspired models \cite{Arkani1998}, strongly suggest that spacetime may possess more than four dimensions. In such theories, the gravitational field and the properties of matter can differ significantly from their four-dimensional counterparts, potentially altering the internal makeup of compact objects. This provides strong motivation to generalize the concept of complexity to higher-dimensional spacetimes, where it may serve as a useful tool for probing how extra dimensions influence the structure and behaviour of self-gravitating systems.

Recently, Choudhury et al. \cite{choudas} extended Herrera's concept of gravitational complexity from the conventional four-dimensional framework to arbitrary $(n+2)$-dimensional static spherically symmetric spacetimes by performing the orthogonal splitting of the Riemann tensor in higher dimensions. Within this generalized formalism, a higher-dimensional scalar complexity factor was derived that simultaneously incorporates the effects of energy-density inhomogeneity and pressure anisotropy while naturally accounting for the dimensional dependence of the gravitational coupling constant and the geometry of the $n$-sphere. The resulting formulation establishes a unified theoretical framework for studying the internal structure of self-gravitating systems in arbitrary spacetime dimensions and provides a natural extension of Herrera's original definition beyond four-dimensional General Relativity.

Although the generalized complexity formalism has now been established, its application to physically realistic compact-star models remains largely unexplored. In particular, exact analytical solutions satisfying the vanishing-complexity condition in higher-dimensional Einstein gravity are still absent from the literature. Constructing such solutions is important because they provide valuable insight into the influence of extra spatial dimensions on the equilibrium, stability, and internal organization of relativistic stellar configurations. Furthermore, exact models serve as useful benchmarks for understanding the role played by higher-dimensional geometry in determining the physical properties of compact stars.

Earlier, several authors obtained complexity-free solutions under different geometric backgrounds \cite{Casadio2019,herrera2020,Maurya2022a,Maurya2022b1,Arias2023,Das2024a,Das2024b} including, most relevantly to the present study, a Finch-Skea dark-energy star closed by the vanishing complexity condition \cite{RejBogadiGovender2024}. All of these constructions, however, remain confined to standard four-dimensional spacetime. Motivated by these considerations, the present work focuses on constructing exact higher-dimensional compact-star models that admit vanishing complexity by adopting the Finch--Skea spacetime \cite{finch1989}. The geometric structure of the Finch--Skea ansatz allows the generalized vanishing-complexity condition to be solved analytically, leading to a new family of exact interior solutions in arbitrary spacetime dimensions. The obtained solutions are examined for physical viability by analysing the behaviour of the matter variables, energy conditions, causality, stability, and equilibrium. These tests confirm that the models satisfy the essential requirements of realistic compact stellar configurations.

An important objective of the present investigation is to understand how the dimensionality of spacetime affects the internal structure of self-gravitating matter under the condition of vanishing complexity. Since both the Einstein field equations and the complexity factor acquire explicit dimensional dependence in higher-dimensional gravity, the physical characteristics of compact stars are expected to differ from those predicted by the conventional four-dimensional theory. Consequently, the present study not only extends the class of exact solutions available in higher-dimensional General Relativity but also provides new insight into the relationship between spacetime geometry, pressure anisotropy, density inhomogeneity, and gravitational complexity.

The present work serves as the first physical application of the generalized higher dimensional complexity factor developed by Choudhury et al. \cite{choudas}, in the context of vanishing complexity factor.

The paper is organized as follows: In Sect.~\ref{section2}, we present the Einstein field equations for static spherically symmetric spacetimes in $(n+2)$ dimensions. Section~3 briefly reviews the orthogonal splitting of the Riemann tensor and the corresponding higher-dimensional complexity factor. In Sect.~4, the vanishing-complexity condition is imposed to obtain the governing differential equation for the metric potentials. Exact Finch--Skea solutions are derived in Sect.~5, while the matching conditions with the higher-dimensional Schwarzschild exterior spacetime are discussed in Sect.~6. The physical properties of the obtained stellar models are analysed in Sect.~7 through the behaviour of the matter variables, anisotropy, energy conditions, causality, stability, and equilibrium. Finally, the principal discussions and conclusions of the present investigation are summarized in Sect.~8 and Sect.~9.

\section{Metric and Field Equations}\label{section2}

We consider a static, spherically symmetric, shear-free spacetime in \((n+2)\)-dimensions, described by the line element
\begin{equation}
    ds^2 = -A_0^2(r) \, dt^2 + B_0^2(r) \, dr^2 + r^2 \, d\Omega_n^2,
    \label{eq:metric}
\end{equation}
where \(A_0(r)\) and \(B_0(r)\) are metric potentials that depend solely on the radial coordinate \(r\). The angular metric \(d\Omega_n^2\) represents the line element on the unit \(n\)-sphere, given by
\[
d\Omega_n^2 = d\theta_1^2 + \sin^2\theta_1 \, d\theta_2^2 + \cdots + \left(\prod_{i=1}^{n-1} \sin^2 \theta_i \right) d\theta_n^2.
\]
This geometry naturally generalizes the familiar four-dimensional spherically symmetric metric to higher spatial dimensions, where the total spacetime dimensionality is \(D = n + 2\).

The matter content is fundamentally modelled starting from the isotropic perfect fluid, whose classical energy-momentum tensor takes the form
\begin{equation}
    T_{ij} = (\rho + p) u_i u_j + p\, g_{ij},
    \label{eq:EM_tensor}
\end{equation}
where \(\rho = \rho(r)\) is the proper energy density, \(p = p(r)\) is the isotropic pressure, and \(u^i = A_0^{-1}(r) \delta_0^i\) is a timelike fluid (n+2) four-velocity normalized such that \(u^i u_i = -1\). However, to accurately model realistic compact stars exhibiting local pressure anisotropy, this isotropic framework is generalized such that the pressure decomposes into a radial component, \(p_r(r)\), and a tangential component, \(p_t(r)\), acting orthogonally to the radial direction.

The Einstein field equations in the \((n+2)\)-dimensional spacetime are governed by
\begin{equation}
    G_{ij} = \kappa_n T_{ij},
\end{equation}
where \(G_{ij}\) is the Einstein tensor computed from the metric \eqref{eq:metric}, and \(\kappa_n\) is the generalized gravitational coupling constant in higher dimensions. Following the conventions adopted by Mansouri and Nayeri \cite{Mansouri}, \(\kappa_n\) is defined as
\begin{equation}
    \kappa_n = \frac{n}{n - 1} \, \frac{2\pi^{\frac{n+1}{2}}}{\Gamma\left(\frac{n+1}{2}\right)},
    \label{eq:kappa_n}
\end{equation}
where \(\Gamma\) denotes the Euler Gamma function. This formulation ensures that both the Newtonian limit and higher-dimensional gravitational extensions remain mathematically consistent.

Evaluating the non-vanishing components of the Einstein tensor for the metric \eqref{eq:metric} and incorporating the anisotropic matter distribution into the field equations yields the following independent relations:

\paragraph{Energy density:}
\begin{equation}
    \kappa_n \rho = \frac{n}{r B_0^2} \left[ \frac{(n-1)}{2r} \left( B_0^2 - 1 \right) + \frac{B_0'}{B_0} \right],
    \label{eq:rho_eq}
\end{equation}
which relates the energy density \(\rho\) to the spatial metric potential \(B_0(r)\) and its radial derivative.

\paragraph{Radial pressure:}
\begin{equation}
    \kappa_n p_r = \frac{n}{r B_0^2} \left[ -\frac{(n-1)}{2r} \left( B_0^2 - 1 \right) + \frac{A_0'}{A_0} \right],
    \label{eq:p_r_eq}
\end{equation}
expressing the radial pressure \(p_r\) in terms of both metric potentials, \(A_0(r)\) and \(B_0(r)\), and the temporal potential's gradient.

\paragraph{Tangential pressure:}
\begin{align}
    \kappa_n p_t = \frac{1}{r B_0^2} \Bigg[ &-\frac{(n-1)(n-2)}{2r} \left( B_0^2 - 1 \right) \nonumber \\
    &+ (n-1) \left( \frac{A_0'}{A_0} - \frac{B_0'}{B_0} \right) + \frac{r B_0}{A_0} \left( \frac{A_0'}{B_0} \right)' \Bigg].
    \label{eq:p_t_eq}
\end{align}
Here, the primes denote ordinary differentiation with respect to the radial coordinate \(r\).


\section{Mathematical Formulation of the Complexity Factor} \label{section3}

Herrera \cite{herrera2018new} established the theoretical foundation for the complexity factor by performing an orthogonal splitting of the Riemann tensor, which yields a set of physically meaningful scalar quantities. One of these fundamental scalars is the complexity factor, denoted as $Y_{TF}$. In a general relativistic fluid, structural complexity can arise from various sources, including pressure anisotropy, energy-density inhomogeneities, and dissipative heat fluxes. Since the metric considered here represents a static, spherically symmetric spacetime, dissipative energy fluxes are absent (i.e., $T_{tr} = T_{rt} = 0$).

The Weyl tensor decomposes into an electric part and a magnetic part. However, the magnetic part, conventionally denoted by $Z_{\alpha \beta}$, identically vanishes under the assumption of spherical symmetry \cite{frame_drag}. Consequently, the orthogonal decomposition of the Riemann tensor involves only the electric part of the Weyl tensor, parameterized by $E_{\alpha \beta}$, and the structure scalars $X_{\alpha \beta}$ and $Y_{\alpha \beta}$.

The electric part of the Weyl tensor is defined as:
\begin{equation}
    E_{\alpha \beta} = C_{\alpha \mu \beta \nu} u^\mu u^\nu
\end{equation}
where $u^\mu$ denotes the fluid four-velocity defined previously. This definition implies the following non-vanishing components:
\begin{equation}
    E_{11} = \frac{2}{3} B_0^2 \varepsilon
\end{equation}
\begin{equation}
    E_{22} = - \frac{1}{3} r^2 \varepsilon
\end{equation}
\begin{equation}
    E_{mm} = E_{(m-1)(m-1)} \sin ^2 \theta_{m-2}\,\, , \,\,\,\,\,m=3,4,...n+2.\,\,\,\,
\end{equation}

In the above expressions, the scalar $\varepsilon$ is defined in terms of the metric potentials as:
\begin{align}
    \varepsilon = \frac{1}{2B_0^2} \left[  \frac{A_0''}{A_0} + \left(  \frac{B_0'}{B_0} + \frac{1}{r}  \right)   \left(  \frac{1}{r} - \frac{A_0'}{A_0} \right)  \right]  -  \frac{1}{2\,r^2}
\end{align}

The tensor components can be written in a compact, trace-free form as:
\begin{equation}\label{electric}
    E_{\alpha \beta} = \varepsilon\left( \chi_\alpha \chi _\beta - \frac{1}{3} h_{\alpha \beta}  \right) 
\end{equation}
where $\chi_\alpha$ is a unit spacelike four-vector aligned with the radial direction, and $h_{\alpha \beta} = g_{\alpha \beta} + u_\alpha u_\beta$ acts as the projection operator onto the spatial hypersurface orthogonal to the four-velocity, effectively eliminating the temporal components from the calculation.

Next, we decompose the tensors $X_{\alpha \beta}$ and $Y_{\alpha \beta}$ into their trace and trace-free scalar components, denoted as $X_T, X_{TF}, Y_T,$ and $Y_{TF}$, through the following expressions:
\begin{equation}
    Y_{\alpha \beta} = \frac{1}{3} Y_T h_{\alpha \beta} + Y_{TF} \left(\chi_\alpha \chi _\beta - \frac{1}{3}h_{\alpha \beta}\right)
\end{equation}
\begin{equation}
    X_{\alpha \beta} = \frac{1}{3} X_T h_{\alpha \beta} + X_{TF} \left(\chi_\alpha \chi _\beta - \frac{1}{3}h_{\alpha \beta}\right)
\end{equation}

By utilizing the higher-dimensional Einstein field equations \eqref{eq:rho_eq}--\eqref{eq:p_t_eq} along with Eq.~\eqref{electric}, and defining the local pressure anisotropy as $\Pi \equiv p_r - p_t$, we obtain the following scalar relations following the framework established by Herrera \cite{herrera2020}:
\begin{equation}
    Y_T = \frac{n-1}{n}\kappa_n\left( \rho + 3 p_r - 2\Pi  \right),\quad Y_{TF} = \varepsilon - \frac{n-1}{n}\kappa_n \Pi,
\end{equation}
\begin{equation}
    X_T = \kappa_n \rho, \quad X_{TF} = -\varepsilon - \frac{n-1}{n}\kappa_n \Pi. 
\end{equation}

The scalar $Y_{TF}$ corresponds precisely to the complexity factor defined by Herrera. Therefore, we generalize Herrera's notion of gravitational complexity to higher-dimensional spacetimes through the expression:
\begin{equation}
    Y_{TF} = \varepsilon - \frac{n-1}{n}\kappa_n \Pi 
\end{equation}

To express this complexity factor in a more physically transparent form, we recall that the mass function for a homogeneous $n$-dimensional configuration is given by:
\begin{equation}
    m(r) = \int_0^r\rho(s).\,\frac{2\pi^{(n+1)/2}}{\Gamma[(n+1)/2]} s^{n} ds
\end{equation}
Integrating this expression by parts yields:
\begin{equation}
    m = \frac{2\pi^{(n+1)/2}}{\Gamma(\frac{n+1}{2})} \left[   \rho \frac{r^{n+1}}{n+1} - \int_0^r \rho'.\frac{s^{n+1}}{n+1} ds \right]
\end{equation}
\begin{equation}
    \implies \frac{(n+1)m}{r^{n+1}} = \frac{n-1}{n} \kappa_n \left[ \rho - \frac{1}{r^{n+1}}\int_0^r\rho ' s^{n+1} ds   \right]
\end{equation}

Through straightforward algebraic manipulation, we can recast the above equation into the following form:
\begin{equation}
    \frac{(n+1)m}{r^{n+1}} = \frac{n-1}{n}\kappa_n \left( \rho - \Pi  \right) -\varepsilon
\end{equation}
This directly simplifies the expression for the higher-dimensional complexity factor, yielding:
\begin{equation} \label{complexity}
     Y_{TF} = -\frac{2(n-1)}{n}\kappa_n \Pi +\frac{(n-1)\kappa_n}{n .r^{n+1}}. \int_0^r\rho'(s).\,s^{n+1} ds 
\end{equation}

By explicitly substituting the metric variables using the Einstein field equations, we arrive at the following expression for the complexity factor tailored to our chosen geometry:
\begin{align}
    Y_{TF} = \frac{1}{n B_0^2} \Bigg[ & \frac{(2-n)(1-B_0^2)}{r^2} + \frac{2 A_0''}{A_0}  \nonumber \\
    &- \frac{2 A_0'}{r A_0} - \frac{2 A_0' B_0'}{A_0 B_0} + \frac{(2-n) B_0'}{r B_0} \Bigg]. 
\end{align}
This particular generalized result was previously derived in the work by Choudhury \emph{et al.} \cite{choudas}.

We can conceptually separate the complexity factor $Y_{TF}$ defined in Eq.~\eqref{complexity} into two distinct physical contributions:
\begin{equation}
Y_{TF} = Z_1+ Z_2,    
\end{equation}
where $Z_1$ represents the contribution due to local pressure anisotropy, defined as:
\begin{equation}
Z_1=-2\frac{(n-1)}{n}\kappa_n(p_{r}-p_{t}),    
\end{equation}
and $Z_2$ represents the contribution arising from the energy-density inhomogeneity, given by:
\begin{equation}
Z_2=\frac{(n-1)\kappa_n}{n .r^{n+1}}. \int_0^r\rho'(s).\,s^{n+1} ds. \label{Z2}   
\end{equation}
Under the specific condition of vanishing complexity ($Y_{TF} = 0$), there is no need to calculate the term $Z_2$ independently, as the precise balance between density gradients and anisotropy dictates:
\begin{equation}
Z_2=-Z_1=2\frac{(n-1)}{n}\kappa_n(p_{r}-p_{t}). \label{eq:Z2_balance}   
\end{equation}

Finally, for the specific case of $n=2$, the expression naturally reduces to Herrera's original complexity definition for four-dimensional spacetime:
\begin{equation}
    Y_{TF} = \frac{1}{B_0^2} \left[  \frac{A_0''}{A_0} - \frac{A_0'}{r A_0} - \frac{A_0' B_0'}{A_0 B_0}  \right]
\end{equation}

\section{Vanishing Complexity} \label{section4}  

The complexity factor naturally vanishes if the system is both homogeneous and isotropic, in which case both terms on the right-hand side of Eq.~\eqref{complexity} evaluate to zero. Alternatively, a zero-complexity scenario arises when the local pressure anisotropy and the energy-density gradient exactly balance each other, satisfying the relation:
\begin{equation}
    \Pi(r) =\frac{1}{2 r^{n+1}} \int_0^r \rho'(s) \, s^{n+1} ds.
\end{equation}

Substituting the metric variables into the condition $Y_{TF} = 0$, we obtain the following governing equation for a system with vanishing complexity:
\begin{align}
    \Bigg[ & \frac{(2-n)(1-B_0^2)}{r^2} + \frac{2A_0''}{A_0}  \nonumber \\
    &- \frac{2A_0'}{rA_0} - \frac{2A_0' B_0'}{A_0 B_0} + \frac{(2-n)B_0'}{r B_0} \Bigg]=0. \label{vanish_complx}
\end{align}

This expression can be rewritten as a second-order linear differential equation for the temporal metric potential $A_0(r)$:
\begin{align}
    A_0''(r)- f(r) A_0'(r) + g(r) A_0(r)=0,
\end{align}
where the coefficient functions $f(r)$ and $g(r)$ are defined as:
\begin{align}
    f(r)=\left(\frac{1}{r}+\frac{B_0'(r)}{B_0(r)}\right) 
\end{align}
and 
\begin{align}
    g(r)=\left(\frac{2-n}{2r}\right)\left(\frac{B_0'(r)}{B_0(r)}+\frac{1-B_0^2(r)}{r}\right). 
\end{align}
The general solution to this differential equation can be expressed as:
\begin{align}
    A_0(r)
    &= e^{\frac{1}{2}\int f(r)\,dr}
    \left[
    C_1\,y_1(r)
    + C_2\,y_1(r)\int_{0}^{r}\frac{ds}{y_1(s)^2}
    \right],
    \\[6pt]
    \text{where } \quad
    y_1''(r)
    &+ \left(
    g(r)
    + \frac{1}{2}f'(r)
    - \frac{1}{4}f(r)^2
    \right)y_1(r) = 0.
\end{align}

Alternatively, one can recast the vanishing-complexity condition as a Bernoulli differential equation for the spatial metric potential $B_0(r)$:
\begin{align}
    B_0'(r)- F(r) B_0^3(r) - G(r) B_0(r)=0,
\end{align}
where the new coefficient functions $F(r)$ and $G(r)$ are defined in terms of $A_0(r)$ as:
\begin{align}
    F(r) &=
    \frac{n-2}
    {r^2\left(\dfrac{2A_0'}{A_0}-\dfrac{2-n}{r}\right)},
    \\[6pt]
    G(r) &=
    \frac{
    \dfrac{2A_0''}{A_0}
    - \dfrac{2A_0'}{rA_0}
    + \dfrac{2-n}{r^2}
    }{
    \dfrac{2A_0'}{A_0}
    - \dfrac{2-n}{r}
    }.
\end{align}
This nonlinear equation admits the exact solution:
\begin{align}
    B_0(r)
    &=
    \left[
    \frac{
    e^{\displaystyle \int G(r)\,dr}
    }{
    \sqrt{
    C_0 - 2\displaystyle\int F(r)\,e^{\displaystyle 2\int G(r)\,dr}\,dr
    }
    }
    \right],
\end{align}
where $C_1$, $C_2$, and $C_0$ are constants of integration.

\section{Finch--Skea Background Geometry} \label{section5}

Motivated by higher-dimensional extensions of the Finch--Skea model \cite{finch1989}, we assume the spatial metric function $B_0(r)$ takes the form
\begin{equation}
    B_0(r) = \left(1 + C r^2 \right)^{1/2},
\end{equation}
where the parameter $C$ characterizes the spatial curvature of the stellar interior. 

Substituting this metric potential into the vanishing-complexity condition (from Eq. \eqref{vanish_complx}) yields:
\[
    \frac{A_0''}{A_0}
    -\frac{A_0'}{rA_0}
    -\frac{Cr}{(1+Cr^2)}\frac{A_0'}{A_0}
    -\frac{(2-n)C^2 r^2}{2(1+Cr^2)}
    =0.
\]
Upon simplification, we obtain:
\[
    A_0''
    -\frac{(1+2 C r^2)}{r(1+C r^2)}A_0'
    +\frac{(n-2)C^2 r^2}{2(1+Cr^2)}A_0
    =0.
\]

To solve this differential equation, we introduce the coordinate transformation $x = C r^2$ and define $A_0(r) \equiv y(x)$. The first and second derivatives with respect to $r$ then become:
\[
    \frac{d A_0}{dr} = 2\sqrt{xC} \, \frac{dy}{dx},
\]
\[
    \frac{d^2 A_0}{dr^2} = 2C \frac{dy}{dx} + 4Cx \frac{d^2y}{dx^2}.
\]
Substituting these expressions back into the simplified differential equation gives:
\[
    4xC \frac{d^2y}{dx^2} - \frac{2xC}{(1+x)} \frac{dy}{dx} + \frac{n-2}{2}\frac{xC}{(1+x)} y(x) = 0.
\]
Multiplying both sides by $\frac{x+1}{Cx}$, we arrive at:
\[
    2(1+x) \frac{d^2y}{dx^2} - \frac{dy}{dx} + \frac{n-2}{4} y(x) = 0.
\]
Applying a subsequent substitution, $z = 1+x$, the equation transforms into:
\[
    2z \frac{d^2 y}{dz^2} - \frac{dy}{dz} + \frac{n-2}{4} y(z) = 0.
\]
Finally, introducing the variable $t = \sqrt{z}$ yields:
\[
    \frac{d^2 y}{dt^2} - \frac{2}{t} \frac{dy}{dt} + \lambda^2 y(t) = 0,
\]
where $\lambda^2 = \frac{n-2}{2}$. This represents a spherical Bessel-type differential equation. The general solution is:
\begin{equation}
    A_0 = C_1 \left[ \sin(\lambda t) - \lambda t \cos(\lambda t) \right] 
    + C_2 \left[ \cos(\lambda t) + \lambda t \sin(\lambda t) \right],
\end{equation} 
where $C_1$ and $C_2$ are constants of integration, and
\begin{equation}
    \lambda = \sqrt{\frac{n - 2}{2}}, \quad 
    t = \sqrt{1 + C r^2}.
\end{equation}

Transforming back to the original radial coordinate $r$, the temporal metric potential becomes:
\begin{eqnarray}
    A_0(r) &=& C_1 \Bigg[ \sin\!\left(\sqrt{\frac{(n-2)(1+C r^2)}{2}}\right) \nonumber\\
    && - \sqrt{\frac{(n-2)(1+C r^2)}{2}} 
    \cos\!\left(\sqrt{\frac{(n-2)(1+C r^2)}{2}}\right) \Bigg] \nonumber\\
    && + C_2 \Bigg[ \cos\!\left(\sqrt{\frac{(n-2)(1+C r^2)}{2}}\right) \nonumber\\
    && + \sqrt{\frac{(n-2)(1+C r^2)}{2}}\times \nonumber \\ 
    && \sin\!\left(\sqrt{\frac{(n-2)(1+C r^2)}{2}}\right) \Bigg].
\end{eqnarray}
.
The resulting physical parameters of the stellar configuration are expressed as follows:
\begin{equation}
\kappa_n \rho = \frac{Cn\left(1+n+C(-1+n)r^2\right)}{2\left(1+Cr^2\right)^2}    
\end{equation}

\begin{equation}
\eta(r) \equiv \sqrt{n-2}\,\sqrt{1+Cr^2}
\end{equation}

\begin{eqnarray}
\mathcal{D}(r) & \equiv & \Big(2C_2 - \sqrt{2}\,C_1\,\eta(r)\Big)
\cos\!\left(\frac{\eta(r)}{\sqrt{2}}\right) \nonumber\\
& & {}+ \Big(2C_1 + \sqrt{2}\,C_2\,\eta(r)\Big)
\sin\!\left(\frac{\eta(r)}{\sqrt{2}}\right)
\end{eqnarray}

\begin{eqnarray}
\kappa_n p_r & = & -\frac{Cn}{2(1+Cr^2)\,\mathcal{D}(r)}
\Bigg[\Big(2C_2 - \sqrt{2}\,C_1(n-1)\,\eta(r)\Big) \times \nonumber \\
&& \cos\!\left(\frac{\eta(r)}{\sqrt{2}}\right) 
{}+ \Big(2C_1 + \sqrt{2}\,C_2(n-1)\,\eta(r)\Big)\times \nonumber \\
&& \sin\!\left(\frac{\eta(r)}{\sqrt{2}}\right)\Bigg]
\end{eqnarray}

\begin{eqnarray}
\kappa_n p_t & = & -\frac{Cn}{2(1+Cr^2)^2\,\mathcal{D}(r)}
\Bigg[\Big(2C_2 - \sqrt{2}\,C_1\big(n-1-2Cr^2+ \nonumber \\
&& Cnr^2\big)\,\eta(r)\Big)
\cos\!\left(\frac{\eta(r)}{\sqrt{2}}\right) {}+ \Big(2C_1 + \sqrt{2}\,C_2\big(n-1-\nonumber \\ && 2Cr^2+Cnr^2\big)\,\eta(r)\Big)
\sin\!\left(\frac{\eta(r)}{\sqrt{2}}\right)\Bigg]
\end{eqnarray}
\begin{equation}
Y_{TF} = 0    
\end{equation}
\begin{equation}
\Pi = -\frac{C^2 n r^2}{2(1+Cr^2)^2}    
\end{equation}
\begin{equation}
W(r) = -\frac{C^2 n r^2}{2(1+Cr^2)^2}    
\end{equation}

For the standard four-dimensional case ($n=2$), the differential equation reduces to:
\[
    A_0'' - \left(\frac{1}{r} + \frac{Cr}{1+Cr^2}\right) A_0' = 0.
\]
In this specific scenario, the coefficient of $A_0$ vanishes, fundamentally altering the mathematical structure of the differential equation. Consequently, the general formula derived above is no longer valid, as can be verified by attempting to take the limit $n \to 2$. Hence, the correct four-dimensional Finch--Skea solution must be obtained separately, yielding:
\[
    A_0(r) = C_1(1+Cr^2)^{3/2} + C_2.
\]

\section{Matching conditions} \label{section6}

To ensure physical viability, the interior geometry of the stellar model must smoothly match an appropriate exterior vacuum spacetime at the stellar boundary. For a static, spherically symmetric mass distribution, this exterior region is described by the higher-dimensional Schwarzschild  solution, characterized by a vanishing energy-momentum tensor.

The exterior region of the sphere is described by the line element:
\begin{equation}
ds_{+}^2 = -\left(1-\frac{\mathcal{M}}{r^{n-1}}\right)dt^2 +\left(1-\frac{\mathcal{M}}{r^{n-1}}\right)^{-1} dr^2+ r^2 d\Omega_n^2, \label{Vm}
\end{equation}
where the geometric parameter $\mathcal{M}$ is related to the total physical mass $m$ of the configuration. 

The fundamental Israel junction conditions require the continuity of the metric potentials across the boundary hypersurface, denoted here by $r=R$. Equating the interior Finch--Skea spacetime with the exterior Schwarzschild geometry yields the following matching conditions at the stellar surface:
\begin{align}
A_0^2(r)\Big|_{r=R} &= \left(1-\frac{\mathcal{M}}{R^{n-1}}\right),
\end{align}
\begin{align}
B_0^2(r)\Big|_{r=R} &= \left(1-\frac{\mathcal{M}}{R^{n-1}}\right)^{-1}.
\end{align}

Solving this system of equations evaluated at the boundary provides an explicit expression for the spacetime curvature parameter $C$:
\begin{align}
C &=-\frac{\mathcal{M}}{R \left(\mathcal{M} R - R^n\right)}.
\end{align}

Furthermore, the matching conditions fix the integration constants $C_1$ and $C_2$ introduced in the temporal metric potential. For example, in the specific case of a six-dimensional spacetime ($D=6$, implying $n=4$), the constants take the exact analytical forms:
\begin{equation}
\mathcal{P} \equiv \sqrt{\frac{R^4}{R^4-\mathcal{M} R}} 
\end{equation}

\begin{equation}
C_1 =\frac{1}{4} \left(1-\frac{\mathcal{M}}{R^3}\right)^{3/2} \mathcal{P}
   \Biggl(6 \mathcal{P} \sin \left(\mathcal{P}\right)+2 \cos \left(\mathcal{P}\right)\Biggr)
\end{equation}

\begin{equation}
C_2 =\frac{1}{4} \left(1-\frac{\mathcal{M}}{R^3}\right)^{3/2} \mathcal{P}
   \Biggl(6 \mathcal{P} \cos \left(\mathcal{P}\right)-2 \sin \left(\mathcal{P}\right)\Biggr).
\end{equation}

\section{Physical Analysis}\label{section7}

In this section, we present the physical behavior of the derived exact stellar model for two distinct higher-dimensional configurations viz. a five-dimensional spacetime ($n=3$) and a six-dimensional spacetime ($n=4$). Furthermore, we investigate the radial profile of the generalized complexity factor to verify the vanishing-complexity condition across different dimensions.

To establish the physical viability \cite{Ivanov2017}-\cite{SuarezUrango2023} and mathematical consistency of the proposed higher-dimensional compact star model, several standard acceptability conditions are imposed on the interior solution. The metric potentials and matter variables, namely the energy density $\rho$, radial pressure $p_r$, and tangential pressure $p_t$, must remain finite, continuous, and regular throughout the stellar interior, particularly at the center, thereby avoiding any central singularity. The energy density is required to be non-negative and to attain its maximum value at the core, decreasing monotonically towards the stellar surface. Similarly, both pressure components should be positive within the star and decrease outward, with the radial pressure satisfying the boundary condition $p_r(R)=0$ at the stellar surface. In addition, the interior geometry must be smoothly matched to the appropriate exterior vacuum spacetime at the boundary through the corresponding junction conditions. Spherical symmetry further demands the vanishing of anisotropic stress at the center, i.e., $\Delta(0)=0$. The physical nature of the matter distribution is examined through the standard null, weak, strong, and dominant energy conditions \cite{HawkingEllis1973}-\cite{Kolassis1988} which are expressed as 
\begin{align*}
E_1 &= \rho + p_r + 2p_t \geq 0,\\
E_2 &= \rho - p_r \geq 0,\\
E_3 &= \rho - p_t \geq 0.
\end{align*}
Causality \cite{Zeldovich1962} is ensured by requiring the radial and tangential sound speeds to remain within the allowed range which are expressed as $0\leq dp_r/d\rho\leq1$ and $0\leq dp_t/d\rho\leq1$. For stelllar stability, the cracking conditions \cite{Herrera1992}-\cite{Abreu2007}, which is essentially the difference between the squared tangential and radial sound speeds, must satisfy
\begin{equation*}
0 < v_r^2-v_t^2 < 1,
\label{eq:cracking}
\end{equation*}
where
\begin{equation*}
v_r^2=\frac{d p_r}{d\rho},
\qquad
v_t^2=\frac{d p_t}{d\rho}.
\end{equation*}

The classical adiabatic index $\Gamma>4/3$ is required for a stable stellar structure \cite{Chandrasekhar1964}-\cite{Bondi1964}. 
For relativistic, anisotropic stars the condition for stability is given \cite{Heintzmann1975}-\cite{ChanHerreraSantos1993} by:
\begin{equation}\label{stab_corr}
\Gamma > \frac{4}{3} + \left[ -\frac{4}{3}\frac{(p_{r}-p_{t})}{|p'_{r}|\,r} + \frac{1}{3}8\pi\frac{\rho p_{r}}{|p'_{r}|}\,r \right]_{\max}.    
\end{equation}
It is evident from the right middle term of (\ref{stab_corr}) that $P_r-P_t>0$ renders the fluid distribution more stable against collapse, while the opposite is true for $P_r-P_t<0$. The last term (positive in sign) is the relativistic correction to the stability and makes the distribution more susceptible
to collapse by increasing $\Gamma$.

The equilibrium of an anisotropic compact star is governed by the balance among gravitational, hydrostatic, and anisotropic forces. The gravitational force $F_g$ acts inward, while the hydrostatic force $F_h$ counteracts gravity through the pressure gradient. The anisotropic force $F_a$ arises from the difference between the radial and tangential pressures and may either support or oppose the gravitational pull. These forces are expressed as
\begin{equation*}
f_g=-(\rho+p_r)\frac{A'_0(r)}{A_0(r)},\qquad
f_h=-\frac{dp_r}{dr},\qquad
f_a=\frac{2}{r}(p_t-p_r).
\end{equation*}
For a physically viable configuration, these forces must remain in equilibrium throughout the stellar interior, satisfying the generalized \cite{PoncedeLeon1987}-\cite{Riazi2016} Tolman-Oppenheimer-Volkoff (TOV) equation
\begin{equation*}
f_g+f_h+f_a=0.
\end{equation*}
Collectively, these criteria provide a comprehensive assessment of the regularity, physical acceptability, causality, boundary consistency, and stability of the proposed compact star solutions.

\begin{figure}[H]
 \centering
 \includegraphics[scale=0.80]{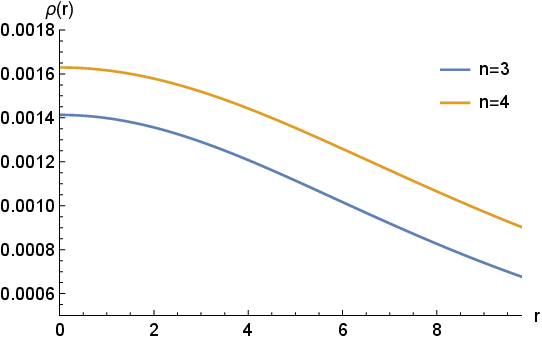}
 \caption{Energy density}
 \label{fig1}
\end{figure}

\begin{figure}[H]
 \centering
 \includegraphics[scale=0.80]{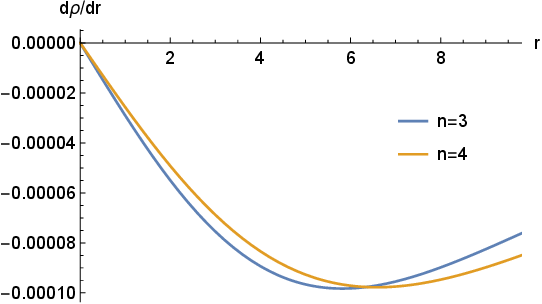}
 \caption{Energy density gradient}
 \label{fig2}
\end{figure}

\begin{figure}[H]
 \centering
 \includegraphics[scale=0.80]{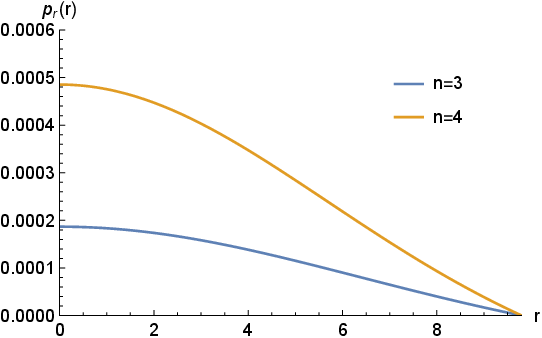}
 \caption{Radial pressure}
 \label{fig3}
\end{figure}

\begin{figure}[H]
 \centering
 \includegraphics[scale=0.80]{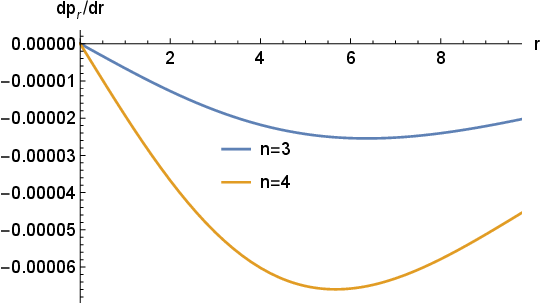}
 \caption{Radial pressure gradient}
 \label{fig4}
\end{figure}

\begin{figure}[H]
\centering
\includegraphics[scale=0.80]{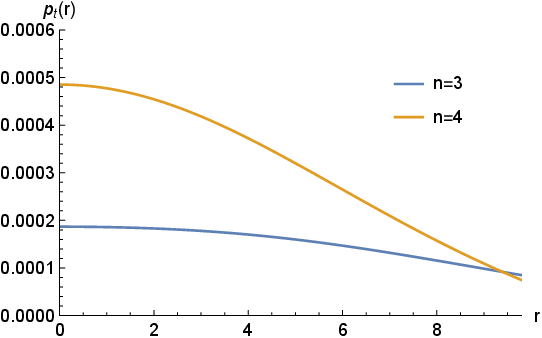}
\caption{Tangential pressure}
\label{fig5}
\end{figure}

\begin{figure}[H]
 \centering
 \includegraphics[scale=0.80]{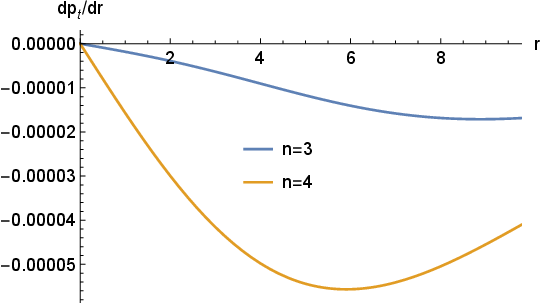}
 \caption{Tangential pressure gradient}
 \label{fig6}
\end{figure}

\begin{figure}[H]
 \centering
 \includegraphics[scale=0.80]{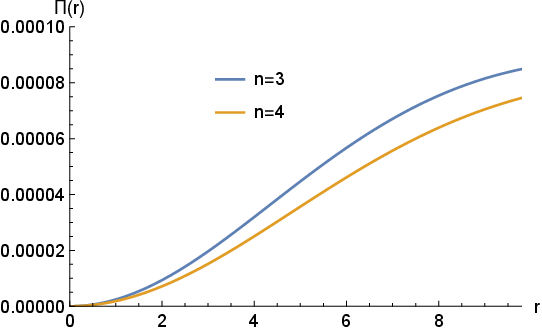}
 \caption{Pressure anisotropy, $\Pi = p_t-p_r$}
 \label{fig7}
\end{figure}

\begin{figure}[H]
 \centering
 \includegraphics[scale=0.80]{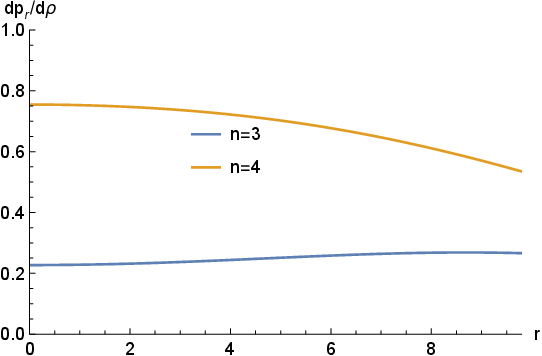}
 \caption{Variation of the radial sound speed}
 \label{fig8}
\end{figure}

\begin{figure}[H]
 \centering
 \includegraphics[scale=0.80]{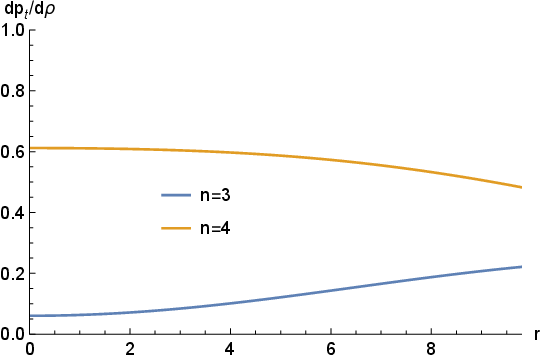}
 \caption{Variation of the tangential sound speed}
 \label{fig9}
\end{figure}

\begin{figure}[H]
 \centering
 \includegraphics[scale=0.80]{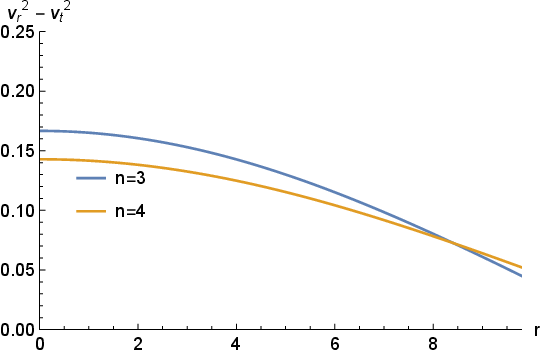}
 \caption{Cracking stability criterion, $v_r^2 - v_t^2$}
 \label{fig10}
\end{figure}

\begin{figure}[H]
 \centering
 \includegraphics[scale=0.8]{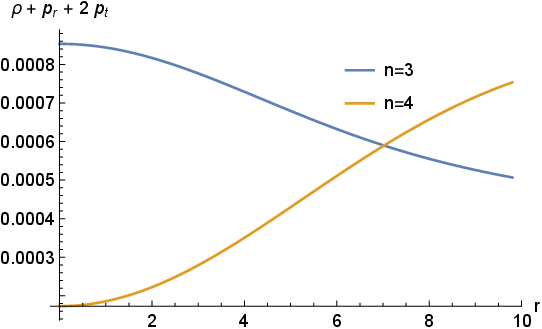}
 \caption{Energy condition $E_1$ }
 \label{fig11}
\end{figure}

\begin{figure}[H]
 \centering
 \includegraphics[scale=0.8]{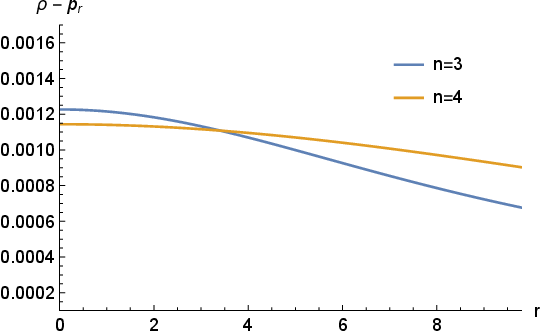}
 \caption{Energy condition $E_2$ }
 \label{fig12}
\end{figure}

\begin{figure}[H]
 \centering
 \includegraphics[scale=0.8]{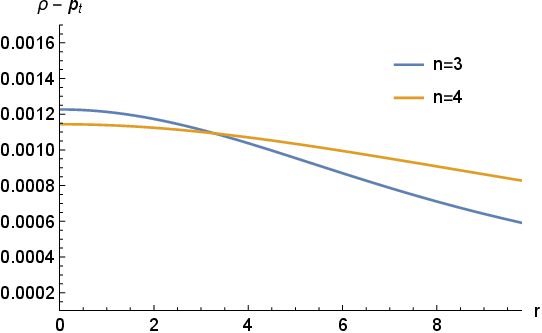}
 \caption{Energy condition $E_3$ }
 \label{fig13}
\end{figure}

\begin{figure}[H]
 \centering
 \includegraphics[scale=0.80]{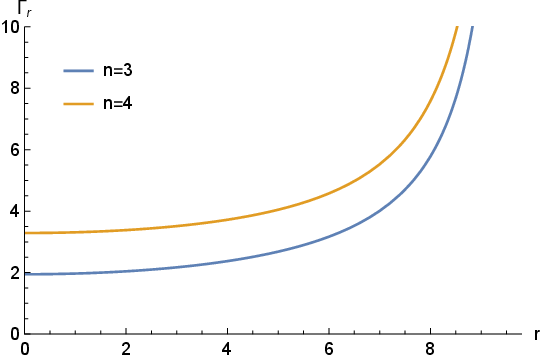}
 \caption{Profile of the radial adiabatic index}
 \label{fig14}
\end{figure}

\begin{figure}[H]
 \centering
 \includegraphics[scale=0.80]{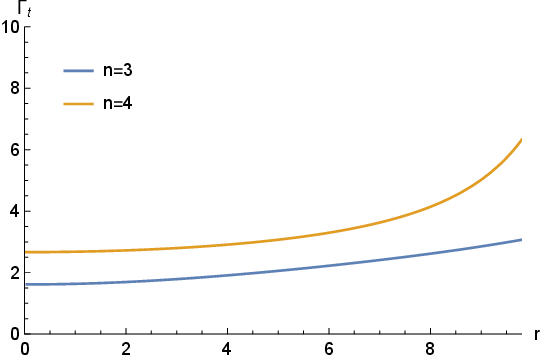}
 \caption{Profile of the tangential adiabatic inde}
 \label{fig15}
\end{figure}

\begin{figure}[H]
 \centering
 \includegraphics[scale=0.80]{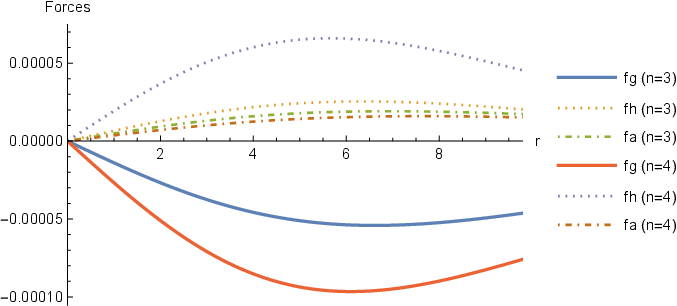}
 \caption{Hydrostatic equilibrium forces (gravitational, hydrostatic, and anisotropic) acting on the system.}
 \label{fig16}
\end{figure}

The physical behaviour of the obtained compact-star configurations is illustrated in Figs.~\ref{fig1}--\ref{fig16} for the representative cases ($n=3$) and ($n=4$). As shown in Fig.~\ref{fig1}, the energy density is positive and finite at the centre and decreases smoothly with increasing radius. Its derivative, plotted in Fig.~\ref{fig2}, remains negative, confirming the monotonic decrease of density. Similar behaviour is obtained for the radial and tangential pressures in Figs.~\ref{fig3} and~\ref{fig5}, while their gradients in Figs.~\ref{fig4} and~\ref{fig6} are negative throughout the stellar interior. These results indicate that the matter variables remain regular and well behaved.

The anisotropy profile in Fig.~\ref{fig7} is non-zero and varies smoothly with radius and vanishes at the centre ($r=0$), showing that the configurations are anisotropic. The radial and tangential sound speeds displayed in Figs.~\ref{fig8} and~\ref{fig9} remain within the causal limit, ($0\leq v_r^2,~v_t^2\leq1$). The cracking criterion in Fig.~\ref{fig10} is consistent with stability. Furthermore, the energy conditions shown in Figs.~\ref{fig11}--\ref{fig13} remain satisfied, while the radial and tangential adiabatic indices in Figs.~\ref{fig14} and~\ref{fig15} support dynamical stability. The force profiles in Fig.~\ref{fig16} show the balance among the gravitational, hydrostatic, and anisotropic forces, thereby confirming hydrostatic equilibrium.

The dimensional parameter ($n$) has a clear influence on these profiles. Since the higher-dimensional field equations and the gravitational coupling explicitly depend on ($n$), increasing the number of spatial dimensions modifies the internal distribution of density, pressure, and anisotropic stresses. In particular, the profiles become more pronounced with increasing ($n$), indicating stronger dimensional effects on the stellar structure. Thus, the comparison between ($n=3$) and ($n=4$) demonstrates that spacetime dimensionality is not merely a mathematical extension, but directly affects the physical properties and equilibrium of the compact star.

\begin{figure}[H] \centering
 \includegraphics[scale=0.80]{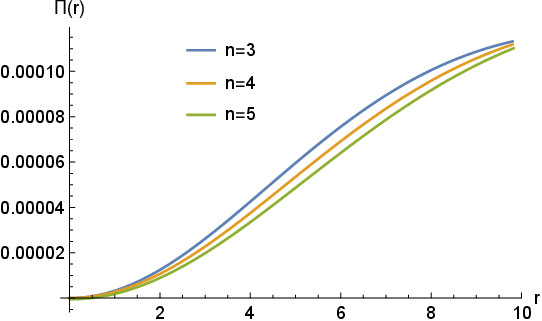}
\caption{Radial profiles of $\Pi(r)$ for $n=3,4,5$.}
 \label{fig7y}
\end{figure}

\begin{figure}[H]
 \centering
 \includegraphics[scale=0.80]{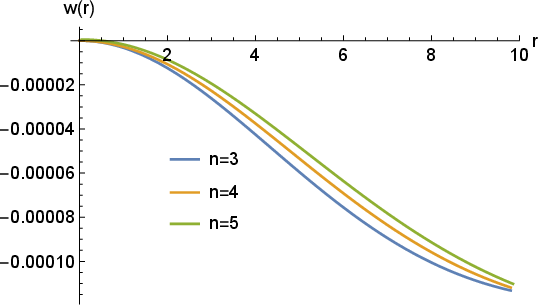}
 \caption{Radial profile of $w(r)$  for $n=3,4,5$.}
 \label{fig8y}
\end{figure}

In Figs.~\ref{fig7y}--\ref{fig8y}, we illustrate the behavior of the two constituent contributions to the complexity factor, namely the pressure anisotropy $\Pi(r)$ and the density inhomogeneity $w(r)$, for different spacetime dimensions. The results indicate that both $\Pi(r)$ and $w(r)$ decrease progressively with increasing dimensionality. Despite this dimensional dependence, their contributions exactly compensate each other, leading to an identically vanishing generalized complexity factor, $Y_{\mathrm{TF}}(r)=0$, throughout the stellar interior. This exact cancellation is consistently maintained for all the considered dimensions, $n=3,4,$ and $5$, demonstrating that the proposed configurations correspond to zero-complexity stellar structures.




\section{Discussion}\label{section8}

The present investigation provides a robust analytical realization of the newly developed higher-dimensional gravitational complexity formalism applied to compact stellar configurations. By coupling the generalized vanishing-complexity condition with the regular, well-behaved Finch--Skea geometric background, we have derived an exact, non-singular family of interior solutions valid in arbitrary $(n+2)$-dimensional Einstein gravity. While many higher-dimensional stellar models rely heavily on numerical integration or the assumption of ad hoc equations of state, our solutions are entirely analytical. This mathematical transparency allows for a rigorous examination of how spacetime dimensionality fundamentally alters the internal structural properties of compact objects.

A central feature of this model is the profound physical implication of the vanishing-complexity condition ($Y_{TF} = 0$). Our analysis unequivocally demonstrates that zero complexity does not merely imply a trivial, homogeneous, or isotropic stellar configuration. Instead, it dictates a state of delicate internal equilibrium where the non-zero density gradient and the local pressure anisotropy ($\Pi = p_r - p_t$) precisely counterbalance one another. As visually confirmed in our complexity profiles (Sect.~\ref{section7}), the contribution arising from the mass-density inhomogeneity diverges in exact opposition to the anisotropic contribution across the radial coordinate, maintaining a net zero complexity factor. This elegantly supports Herrera's paradigm: gravitational complexity is intrinsically linked to the interplay and cancellation of distinct internal fluid properties rather than the absolute absence of those properties.

The physical viability of the derived models has been comprehensively validated. The Finch--Skea spatial potential, $B_0(r) = (1+Cr^2)^{1/2}$, inherently guarantees geometric regularity at the stellar core, while our strict application of the Israel junction conditions seamlessly matches the interior fluid to the  higher-dimensional Schwarzschild exterior  vacuum at the boundary surface. The matter variables---energy density and radial/tangential pressures---remain finite, strictly positive, and monotonically decreasing towards the surface, ensuring that all standard energy conditions are satisfied. The outward-pointing anisotropic force provides crucial additional support against gravitational collapse. Furthermore, the configurations are dynamically stable: the sound speeds respect the causality limits ($0 \leq v_r^2, v_t^2 \leq 1$), Herrera's cracking criterion confirms stability against local perturbations, and the generalized Tolman--Oppenheimer--Volkoff equation establishes a perfect equilibrium among gravitational, hydrostatic, and anisotropic forces. 

A primary objective of this work was to evaluate the footprint of extra spatial dimensions on stellar architecture. Because both the generalized complexity factor and the Einstein field equations depend explicitly on $n$ (via the gravitational coupling $\kappa_n$ and the hyperspherical geometry), increasing the dimensionality fundamentally re-weights the relationship between spacetime geometry and matter. Our empirical plots demonstrate that as $n$ increases, the internal density distribution, pressure profiles, and required anisotropic stresses steepen. 

Consequently, an important physical consequence emerges: viable stellar configurations under vanishing complexity exist only within a highly restricted range of spacetime dimensions. While the differential equations mathematically admit solutions for arbitrary $n$, the stringent physical acceptability criteria---particularly the positivity of pressure, subluminal sound speeds, and cracking stability---become increasingly difficult to satisfy in higher dimensions. Beyond a critical dimensionality threshold, the solutions inevitably violate one or more physical constraints. Thus, the allowable dimensions are strictly filtered by the combined requirements of thermodynamic regularity and hydrostatic stability.

Ultimately, this study bridges exact analytical General Relativity with the modern concept of gravitational complexity. By extending the highly successful Finch--Skea geometry to higher-dimensional zero-complexity systems, we have established a physically sound and theoretically rigorous framework. These exact solutions stand as essential analytical benchmarks for future numerical studies of compact stars in modified theories of gravity, braneworld scenarios, and string-inspired models where higher-dimensional effects cannot be ignored.

\section{Conclusion}\label{section9}

The primary objective of this work was to construct and analyze exact compact-star models that satisfy the generalized vanishing-complexity condition in arbitrary $(n+2)$-dimensional Einstein gravity. By employing the Finch--Skea spacetime as the underlying geometric background, we obtained a new class of exact analytical solutions. To the best of our knowledge, this constitutes the first explicit realization of the recently proposed higher-dimensional complexity formalism within a physically acceptable stellar model.

The generalized complexity condition was successfully incorporated into the higher-dimensional Einstein field equations, yielding exact analytical expressions for the metric potentials and thermodynamic variables. The resulting configurations satisfy all essential physical requirements expected of realistic compact stars. These include geometric regularity at the stellar core, strictly positive and monotonically decreasing matter variables, fulfilment of the standard energy conditions, causal sound propagation, and hydrostatic equilibrium via the generalized TOV equation. Furthermore, the decomposition of the complexity factor definitively shows that a zero-complexity state ($Y_{TF}=0$) is not a trivial uniform fluid distribution; rather, it emerges from a precise, dynamic compensation between the mass-density inhomogeneity and the local pressure anisotropy.

One of the most significant outcomes of this investigation is identifying the profound role that spacetime dimensionality plays in determining stellar viability. Our detailed physical analysis of specific higher-dimensional configurations (e.g., $n=3$ and $n=4$) confirmed that these zero-complexity models maintain dynamical stability against local perturbations, successfully satisfying the cracking criterion. However, we demonstrated that increasing the dimensionality introduces progressively stronger constraints on the internal equilibrium. Consequently, while exact analytical solutions exist mathematically for arbitrary $n$, physically admissible models are restricted to a limited range of dimensions. The physical requirements of causality, positive pressure, and stability act as strict natural filters for higher-dimensional stellar structures.

These solutions extend the highly successful Finch--Skea model beyond the conventional four-dimensional framework, establishing essential analytical benchmark models for investigating anisotropic compact objects in theories admitting extra spatial dimensions. The theoretical framework developed here can be extended in several promising directions. Future investigations may incorporate electric charge, slow rotation, dissipative heat fluxes, or exotic equations of state. Moreover, exploring gravitational complexity within alternative higher-curvature theories---such as Lovelock, scalar--tensor, or Einstein--Gauss--Bonnet gravity---could reveal how additional geometric terms generate new forms of structural complexity. 

Ultimately, this work confirms that the generalized concept of gravitational complexity provides a powerful, physically meaningful framework for studying self-gravitating systems across arbitrary dimensions, laying a solid foundation for future relativistic astrophysics research.

\vspace{0.5cm}
\noindent\textbf{Acknowledgements:} S.D. gratefully acknowledges support from the Inter-University Centre for Astronomy and Astrophysics (IUCAA), Pune, India, where part of this work was carried out under its Visiting Research Associateship Programme. S.D. would also like to express gratitude to ICARD, Malda College. MG acknowledges financial support from the National Research Foundation under grant
number 146050. \\
\textbf{Data Availability Statement:} No datasets were created or analyzed during this study.\\
\textbf{Declaration of Competing Interest:} The authors declare that they have no known competing financial interests or personal relationships that could have appeared to influence the work reported in this paper.


\end{document}